\documentclass[aps,prb,amsmath,amssymb,twocolumn,superscriptaddress,showpacs,floatfix]{revtex4-1}

\usepackage{graphicx}
\usepackage{dcolumn}
\usepackage{bm}

\usepackage{amsmath}
\usepackage{amssymb}

\DeclareMathOperator{\epsi}{\varepsilon}

\DeclareMathOperator{\rp}{\mathit r_\mathrm p}
\DeclareMathOperator{\ab}{\mathit a_\mathrm B}

\DeclareMathOperator{\abc}{\mathit a^3_\mathrm B}
\DeclareMathOperator{\nimp}{\mathit n_\mathrm{i}}
\DeclareMathOperator{\Nimp}{\mathit N_\mathrm{p}}

\begin{document}

\title{Magneto-electric effect in doped magnetic ferroelectrics}

\author{O.~G.~Udalov}
\affiliation{Department of Physics and Astronomy, California State University Northridge, Northridge, CA 91330, USA}
\affiliation{Institute for Physics of Microstructures, Russian Academy of Science, Nizhny Novgorod, 603950, Russia}
\author{I.~S.~Beloborodov}
\affiliation{Department of Physics and Astronomy, California State University Northridge, Northridge, CA 91330, USA}

\date{\today}


\begin{abstract}
We propose a model of magneto-electric effect in doped magnetic ferroelectrics. 
This magneto-electric effect does not involve the spin-orbit coupling 
and is based purely on the Coulomb interaction. We calculate magnetic phase 
diagram of doped magnetic ferroelectrics. We show that
magneto-electric coupling is pronounced only for ferroelectrics with low dielectric constant. 
We find that magneto-electric coupling leads to modification of magnetization 
temperature dependence in the vicinity of ferroelectric phase transition. 
A peak of magnetization appears. We find that 
magnetization of doped magnetic ferroelectrics strongly depends on applied electric field.
\end{abstract}

\maketitle
\section{Introduction}

Multiferroic (MF) materials with strongly coupled ferroelectricity and magnetism is an intriguing challenge now days~[\onlinecite{Fiebig2015,Scott2006,Spal2007,Ram2004,Bar2008,Math2008}]. 
Among various MF materials the doped magnetic ferroelectrics (DMFE) attract a lot of 
attention since these materials demonstrate the existence of electric polarization and magnetization at room temperatures~[\onlinecite{Jedrecy2015,Gao2009,Liu2009,Cohn2002,Nan2009,Schmidt2011,Chu2009,Yu2011,Han2009}]. 
DMFEs are fabricated by doping of ferroelecrics (FE) with magnetic impurities. 
Transition metal (TM)-doped BaTiO$_3$ (BTO) 
is the most studied material in this family. While both order parameters are simultaneously 
non-zero in DMFE, the coupling between them (magneto-electric effect) is very weak and not enough studied~[\onlinecite{Chatterjee2017,Du2011,Kotnala2013}]. Mostly the magneto-electric (ME) effect in DMFE is related 
to spin-orbit interaction leading to influence of electric polarization on the material magnetic properties.

In the case of BTO the room temperature ferroelectricity is the internal 
property of the material. Magnetization appears due to artificially introduced magnetic impurities~[\onlinecite{Jedrecy2015,Gao2009,Liu2009,Cohn2002,Nan2009,Schmidt2011,Chu2009,Yu2011,Han2009}]. Several mechanisms of coupling between magnetic impurities are known~[\onlinecite{Fitzgerald2005}]. At high doping 
the adjacent magnetic moments directly interact with each other due to electron wave function overlap. 
This interaction is usually antiferromagnetic. 
At low impurities concentration ($<$10\%) the direct coupling is not possible. 
However, the room temperature ferromagnetic (FM) ordering is observed in this limit. 
The reason for FM interaction between the impurities in this case is shallow donor 
electrons which inevitably present due to defects such as oxygen vacancies. 
Donor electrons have weakly localized wave function spanning over several lattice periods. 
Donor electron interacts with impurities forming so-called bound magnetic polaron (BMP) 
in which all magnetic moments are co-directed. The polaron size essentially 
exceeds the interatomic distance. Interaction of the polarons leads to the 
formation of long-range magnetic order in the system. Due to large BMP size the 
critical concentration of defects and magnetic impurities at which FM ordering appears can be rather low. 
BMPs and their interaction are well understood in doped magnetic 
semiconductors~[\onlinecite{Fitzgerald2005,Spalek1983,Bhatt2001,Wolff2002,Sarma2002}].

In the present work we propose a model of magneto-electric (ME) coupling in DMFE. 
The idea behind this model is based on the fact that 
shallow donor electron interacts not only with magnetic impurities but also with phonons forming not just 
magnetic polaron but the electro-magnetic one. 
Magnetic and orbital degrees of freedom are strongly coupled in such a polaron. In contrast to the 
most magneto-electric effects based on the spin-orbit interaction we 
consider here the ME coupling occurring purely due to the Coulomb interaction. 
Note that the Coulomb based ME effects were considered recently in a number of other systems~[\onlinecite{Bel2014ME1,Bel2014ME2,Beloborodov2017ExGr,Beloborodov2017ExMTJ,Beloborodov2017ExGr1}].

The size of magnetic polaron is defined by the wave function of a donor electron. 
In its turn the size of the donor electron wave function 
is defined by electron-phonon interaction and depends on the dielectric properties of the 
FE matrix~[\onlinecite{Mukhopadhyay1987,Platzman1972,Hattori1975,Whitfield1969}].
Well known that the dielectric constant of FEs strongly depends on temperature 
and applied electric field. This opens a way to control magnetic polarons 
with electric field or temperature. Finally, the magnetization of the whole sample 
becomes dependent on the external parameters. In the present work we study this mechanism 
of ME coupling. In particular, we study magnetic phase diagram of DMFE 
and show that one can control magnetization with electric field in such a system.

In DMFEs based on FEs with high dielectric constant this effect is negligible, 
which is consistent with observed weak ME effect in doped BTO. A good FE matrix 
would be  Hf$_{0.5}$Zr$_{0.5}$O$_2$~[\onlinecite{Mikolajick2011,Frey2011,Mikolajick2012_1}] 
which has a low dielectric constant ($\epsi<50$) strongly dependent 
on applied electric field. Currently there are no data 
on Hf$_{0.5}$Zr$_{0.5}$O$_2$ doped with magnetic impurities. 
Another magnetic FE with low dielectric constant is (Li,TM) co-doped zinc 
oxide~[\onlinecite{Zeng2008,Coey2004,Gehring2008,Waghmare2005}]. The 
ME effect in this material can be also strong.

The paper is organized af follows. We present the model 
in Sec.~\ref{Sec:Model}. Properties of single magnetic and electric polarons 
are discussed in Sec.~\ref{Sec:SinglePolaron}. Mechanisms of interaction of BMPs 
are considered in Sec.~\ref{Sec:TwoPolarons}. Magnetic phase diagram 
of DMFE and ME effect in a number of systems are presented in Sec.~\ref{Sec:Results}.

\section{The model}\label{Sec:Model}

As we mentioned in the Introduction, 
the long range magnetic order in DMFE appears due to interaction of BMPs 
formed by shallow donor electrons and magnetic impurities. 
To understand magnetic properties of DMFE we first study the interaction of two BMPs.

Consider FE with  magnetic impurities localized at points $\mathbf r^\mathrm i_i$. Each impurity has a spin $S_0$. 
The impurities concentration is low (below 20\%~[\onlinecite{Fitzgerald2005}]) 
and there is no direct interaction between them. 
There are also defects with positions $\mathbf r^\mathrm d_i$ in the system. 
Their concentration is smaller than concentration 
of magnetic impurities. Ordinarily, oxygen vacancies serve as such defects. 
A defect creates a point charge potential ($\sim e^2/|\mathbf r-\mathbf r^\mathrm d_i|$). 
A charge carrier is bound to each of these defects. 
The carrier spin is $s_0=1/2$. The carriers (electrons) interact with 
impurity spins forming bound magnetic polarons. Consider two neighbouring defects. 
They are described by the Hamiltonian
\begin{equation}\label{Eq:SPHam}
\hat H=\hat H_\mathrm e +\hat H_\mathrm{ph}+\hat H_\mathrm{e-ph}+\hat H_\mathrm{e-imp},
\end{equation}
where carriers energy is given by
\begin{equation}\label{Eq:DHam}
\hat H_\mathrm e=-\frac{\hbar^2}{2m^*}\sum_{i}\Delta_i-\frac{e^2}{4\pi\epsi_0\epsi}\left(\!\sum_{i,j}\frac{1}{|\mathbf r_i-\mathbf r^\mathrm d_j|}-\frac{1}{|\mathbf r_1-\mathbf r_2|}\!\right)\!.
\end{equation}
Here $m^*$ is the effective mass of electron in conduction band of the material in the model of rigid lattice, $\epsi$ is the static dielectric constant, $\epsi_0$ is the vacuum dielectric constant, $\mathbf r_i$ are the carriers coordinates, indexes 
$i$ and $j$ take two values, $1$ or $2$.

The interaction between the carriers and impurities is given by the Hamiltonian 
\begin{equation}\label{Eq:MHam}
\hat H_\mathrm{e-imp}=J_0 \sum_{i=1,2}\sum_j (\hat{\mathbf{s}}_i\hat{\mathbf{S}}_j)\delta (\mathbf r^\mathrm i_j-\mathbf r_i),
\end{equation}
where $\hat{\mathbf{s}}_i$ and $\hat{\mathbf{S}}_j$ are the electron and impurity spin operators, respectively. 
The impurity spin, $S_0$ is usually much larger than one half. 
$J_0$ is the interaction constant. 
Interaction with magnetic impurities leads to formation of magnetic polaron.

Terms $\hat H_\mathrm{ph}$ and $\hat H_\mathrm{e-ph}$ in Eq.~(\ref{Eq:SPHam}) 
are the Hamiltonians of phonons and electron-phonon interaction, respectively~[\onlinecite{Mukhopadhyay1987}]. 
We assume that carrier interacts mostly with longitudinal optical phonons. Generally, coupling to acoustical phonons and piezoelectric interaction can be taken into account. We neglect them for simplicity, since they are usually 
weaker than interaction with optical phonons and do not lead to any qualitative changes.  The 
electron-phonon coupling leads to formation of electric 
polaron. We will use results of electric polaron theory to describe the electron wave function~[\onlinecite{Mukhopadhyay1987}]. The whole system of electron, magnetic impurities and phonons is an electro-magnetic bound polaron.

\subsection{Dielectric properties of FEs}

Below we will show that dielectric properties of FE matrix play crucial 
role in formation of the magnetic state of DMFE. Therefore, we need to 
introduce some model of dielectric susceptibility for considered FEs. 
For simplicity we assume that dielectric properties of FE matrix are isotropic. 
We introduce the dependence of dielectric permittivity on applied 
electric field below the Curie temperature
\begin{equation}\label{Eq:Diel}
\epsi^{\pm}(E)=\varepsilon_\mathrm{min}^E+\frac{\Delta\varepsilon^E}{1+(E \mp E_\mathrm s)^2/\Delta E_\mathrm s^2}.
\end{equation}
Superscripts ``$+$'' and ``$-$'' correspond to the upper and the
lower hysteresis branch, respectively, $E_\mathrm s$ is the electric field at which the 
electrical polarization switching occurs, $\Delta E_\mathrm s$ is the width of
the switching region. $\varepsilon_\mathrm{min}^E$ and $\Delta\varepsilon^E$ define the 
minimum dielectric constant and its variation with electric field. Equation~(\ref{Eq:Diel}) captures the basic features of dielectric
constant behavior. The permittivity has two branches corresponding to two
polarization states. In the vicinity of the switching field, $E_\mathrm s$ the dielectric permittivity, $\epsi$ has a peak. 

Not much data are currently available on voltage dependencies of $\epsi(E)$ for 
FEs with low dielectric constants. 
The dielectric constant of Hf$_{0.5}$Zr$_{0.5}$O$_2$ can be described using the
following parameters: $\varepsilon^E_\mathrm{min}=30$, $\Delta\varepsilon^E=15$, $E_\mathrm s=0.1$ V/nm, $\Delta E_\mathrm s=0.1$ V/nm.  

We model the temperature dependence of FE dielectric constant using the following formula
\begin{equation}\label{Eq:DielT}
\epsi(T)=\epsi_\mathrm{min}^T+\frac{\Delta\epsi^T}{\sqrt{(T-T^\mathrm{FE}_\mathrm C)^2+\Delta T^2}}.
\end{equation}
This function allows to describe the finite height peak at FE phase transition temperature $T=T^\mathrm{FE}_\mathrm C$ as well as $1/(T-T^\mathrm{FE}_\mathrm C)$ dependence in the 
vicinity of $T=T^\mathrm{FE}_\mathrm C$. For simplicity we neglect different 
behavior of dielectric constant above and below $T^\mathrm{FE}_\mathrm C$. This does not lead to any qualitative changes in the properties of considered system.

\section{Single polaron properties}\label{Sec:SinglePolaron}

First, consider a single electron located at a defect and interacting with impurities and phonons. 
In the models of electric and magnetic polarons the electron wave function is chosen in the form of 
spherically symmetric wave function
\begin{equation}\label{Eq:DWF1}
\Psi=\Psi_0 e^{-r/\ab},
\end{equation}
where $\ab$ is the decay length, $|\Psi_0|^2=1/(\pi\abc)$.

\subsection{Bound magnetic polaron}\label{Sec:SingleMagPol}

First, consider the interaction of bound electron with impurities 
leading to formation of magnetic polaron. Properties of a single magnetic polaron was 
investigated in the past~[\onlinecite{Spalek1983,Wolff1988,Wolff2002,Fitzgerald2005}]. Lets calculate the 
average electron-impurities interaction energy, $\langle \hat H_\mathrm{e-imp}^\mathrm s\rangle$ (superscript $\mathrm s$ 
indicates that we consider a single polaron) for given $\ab$. 
Following Ref.~[\onlinecite{Sarma2002}] we average the magnetic energy over the spatial coordinates 
\begin{equation}\label{Eq:MagHam1}
\hat H_\mathrm{e-imp}^\mathrm s=-\sum_i J_0(\hat{\mathbf{s}}\hat{\mathbf{S}}_i)|\Psi(\mathbf r^\mathrm i_i)|^2.
\end{equation}

The strongest interaction between electron and impurities appears inside the sphere of radius $\rp$
\begin{equation}\label{Eq:PolRad}
\rp=\frac{\ab}{2}\mathrm{ln}\left(\frac{J}{k_\mathrm B T}\right),
\end{equation}
where $J=S_0J_0|\Psi_0|^2/2$, $k_\mathrm B$ is the Boltzman constant, 
and $T$ is the temperature. We assume that $\rp>\ab$ ($J>6k_\mathrm B T$). 
Only in this case magnetic percolation can appear prior to electric percolation 
(insulator metal transition). Number of impurities within the sphere is 
$N_\mathrm p=4\pi r_\mathrm p^3 \nimp/3$, where $\nimp$ is the impurities concentration.

We assume that the number of impurities within the radius $\rp$ is big enough. The 
total spin of impurities relaxes much slower than the spin of the charge carrier. 
We can introduce the ``classical'' exchange field (measured in units of energy) 
acting on the electron magnetic moment
\begin{equation}\label{Eq:PolRad1}
\begin{split}
\mathbf B&=J_0 \langle \sum_i |\Psi(\mathbf r^\mathrm i_i)|^2 \hat {\mathbf S}_i \rangle\approx J_0 \langle\sum_{r^\mathrm i_i<\rp} |\Psi(\mathbf r^\mathrm i_i)|^2  \hat {\mathbf S}_i\rangle\approx \\ &\approx 2 J \langle \sum_{r^\mathrm i_i<\rp} \hat {\mathbf S}_i\rangle/S_0.
\end{split}
\end{equation}
The maximum value of the field is 
$B_\mathrm{max}\approx 2 N_\mathrm p J$. Note than in the absence of 
an electron the impurities spins are independent 
and the average field value is zero. 
Field fluctuations are given by $\sqrt{\langle B^2 \rangle}\approx B_\mathrm{max}/\sqrt{N_\mathrm p}$.

Equation~(\ref{Eq:MagHam1}) can be rewritten as follows
\begin{equation}\label{Eq:MagHam3}
\hat H_\mathrm{e-imp}^\mathrm s=-(\hat{\mathbf{s}}\mathbf B).
\end{equation}
This Hamiltonian has two non-degenerate eigenstates. For $J>k_\mathrm B T$ the average magnitude of the impurities 
field $\sqrt{\langle B^2 \rangle}\gg k_\mathrm B T$ meaning 
that even in the case of independent impurities the electron spin should be correlated with the instant meaning of the average field $\mathbf B$, and $\langle \hat{\mathbf s}\rangle=(1/2) (\mathbf B/B)$. For the energy averaged over the electron spin states we find
\begin{equation}\label{Eq:MagHam4}
H_\mathrm{e-imp}^\mathrm s=\langle \hat H_\mathrm{e-imp}^\mathrm s\rangle=-B/2.
\end{equation}
Now we determine the field $B$ by taking into account the interaction of electron and 
impurities. This interaction does not lead to the appearance of average $\mathbf B$. 
The average absolute value (fluctuations) of $B=\sqrt{\langle \mathbf B^2\rangle}$ 
is non-zero and is defined by the competition of entropy and internal energy. 
To find the average $B$, consider the states of the system close to the state with full polarization of impurities (within sphere $r<\rp$). 
The fully polarized state means that all the impurity spins 
have the same and maximum projection on a certain axis. There is only one such state, but it has the lowest energy.  
If one reduces total impurities spin by 1, the energy increases by $J/S_0$. At the same time the 
number of states with reduced spin is $\Nimp$. If $\mathrm{ln}(\Nimp)\gg J/(S_0k_\mathrm B T)$, the entropy is the stronger factor than internal energy. In this case donor electron can not couple spins of impurities and they are almost independent. 
In this limit $H_\mathrm{e-imp}^\mathrm s\approx -B_\mathrm{max}/(2\sqrt{\Nimp})$. This corresponds to fluctuation regime of BMP. 
In the opposite limit, $\mathrm{ln}(\Nimp)<J/(S_0k_\mathrm B T)$, the 
internal energy is dominant. In this case all impurities spins are correlated due to interaction 
with the electron and $H_\mathrm{e-imp}^\mathrm s=-B_\mathrm{max}/2$. 
For $J=12k_\mathrm B T$ (which is in agreement with our requirement $J_0>6k_\mathrm B T$ 
and corresponds to $\rp \approx 1.25 \ab$) and $S_0=5/2$ we find 
$\mathrm{ln}(\Nimp)<4.8$ and $\Nimp<120$. This estimate is reasonable and the number 
of impurities in BMP is always within this range~[\onlinecite{Fitzgerald2005}]. 
In our work we consider 
the case of well correlated BMP since only in this limit one can expect strong magnetism.

Since $\Nimp\sim \abc$ and $J\sim (\abc)^{-1}$ the magnetic energy of BMP is independent of the characteristic size of the 
wave function, $\ab$. 
Therefore, in this regime the 
interaction with impurities does not influence the electron spatial distribution.

\subsection{Electric polaron}\label{Sec:SingleElPol}

In previous section we have shown that interaction with impurities 
does not influence the Bohr radius of the bound carrier wave function. 
Therefore, $\ab$ is defined by the interaction of the electron with 
defect charge and with phonons. The problem of electric polaron was studied 
in the past~[\onlinecite{Mukhopadhyay1987,Platzman1972,Hattori1975,Whitfield1969}]. 
There are numerous approaches to this problem. We will follow a variational approach 
of Ref.~[\onlinecite{Larsen1969}]. The Hamiltonian of a single electric polaron has the form
\begin{equation}\label{Eq:SinglePolHam}
\hat H_\mathrm p=-\frac{\hbar^2}{2m^*}\Delta-\frac{e^2}{4\pi\epsi_0\epsi r}+\hat H_\mathrm{ph}+\hat H_\mathrm{e-ph}.
\end{equation}
The electron wave function is given in Eq.~(\ref{Eq:DWF1}). Wave function of phonons is given in 
Ref.~[\onlinecite{Larsen1969}]. The radius of electric polaron, $\ab$ is defined by the 
minimization of average energy $\langle\hat H_\mathrm p\rangle$ with respect to $\ab$. 
In the case of strong coupling between the carrier and phonos the Bohr radius is given by~[\onlinecite{Larsen1969}]
\begin{equation}\label{Eq:DecLength}
(\ab)^{-1}=\frac{m^*e^2}{16\hbar^2}\left(\frac{11}{\epsi}+\frac{5}{\epsi_{\infty}}\right),
\end{equation}
where $\epsi_\infty$ is the optical dielectric constant.

In FE materials the static dielectric constant $\epsi$ depends on temperature $T$ and external electric field $E$. Therefore, one can control the donor electron wave function size $a_\mathrm B$ with external electric field or by varying temperature.

For materials with large static dielectric constant $\epsi\sim 1000$ (as in BTO, for example) the 
Bohr radius becomes independent of $\epsi$ ($\ab=16\hbar^2\epsi_{\infty}/(5m^*e^2)$). In this case variation of $\epsi$ with temperature or electric field does not influence the polaron size. 

In a number of FEs the static dielectric constant is of the same order as the optical one. For example, in Hf$_{0.5}$Zr$_{0.5}$O$_2$ the static dielectric constant is about $30$ while optical one is about 4.5 (there is no experimental data 
on $\epsi_{\infty}$ in Hf$_{0.5}$Zr$_{0.5}$O$_2$, therefore we use data on HfO$_2$ and ZrO$_2$ for estimates). 
Static dielectric constant of this material depends on applied electric field~[\onlinecite{Mikolajick2012_1}]. 
According to Eq.~(\ref{Eq:Diel}) the FE dielectric constant has a peak in the 
vicinity of switching field. The polaron radius grows with $\epsi$. Thus, the $\ab(E)$ has also a peak in the 
vicinity of switching field, $E_\mathrm s$. Variation of $\epsi$ with field in Hf$_{0.5}$Zr$_{0.5}$O$_2$ is about 50\% ($\epsi_{\mathrm{min}}^E=30$, $\Delta \epsi^E=15$). This leads to 10\% changes of polaron radius. 

In Li-doped ZnO oxide the static dielectric constant strongly depends on temperature and is not very large. 
FE properties strongly depend on Li concentration. FE phase transition in these materials 
is usually above the room temperature ~[\onlinecite{Kang2009,Krupanidhi2006,Kumar2011,Mehmood2014}]. In the vicinity of the FE Curie temperature the static dielectric constant varies from 5 to 60. Such a strong growth of the dielectric constant can increase the polaron radius twice.

\section{Interaction of two electro-magnetic polarons}\label{Sec:TwoPolarons}

In this section we consider magnetic interaction of two electro-magnetic 
polarons in DMFE. We introduce here magnetic moments of these 
polarons. They have directions $\mathbf m_{1,2}$. 
Since there is a large number of impurities in each polaron we can treat these quantities as classical vectors. The distance $R=|\mathbf r^\mathrm d_1-\mathbf r^\mathrm d_2|$ between these two polarons exceeds $2 a_\mathrm B$ and $2r_\mathrm p$. 
In this case the inter-polaron magnetic interaction is weak comparing to magnetic energy of a 
single polaron. There are three mechanisms of magnetic coupling between polarons: 1) exchange due to the Coulomb interaction in the Hamiltonian Eq.~(\ref{Eq:DHam}) (Heitler-London interaction); 2) magnetic coupling due to kinetic energy term in the Hamiltonian Eq.~(\ref{Eq:DHam}) (superexchange); and 3) magnetic coupling mediated by impurities, Eq.~(\ref{Eq:MHam}).

\subsection{Heitler-London interaction between polarons}

Consider the Hamiltonian in Eq.~(\ref{Eq:DHam}). If two defects 
are far away from each other the Hamiltonian can be split into zero order Hamiltonian of two non-interacting carriers
\begin{equation}\label{Eq:DHamNI}
\hat H^{(0)}_\mathrm e=-\frac{\hbar^2}{2m^*}\sum_{i}\Delta_i-\frac{e^2}{4\pi\epsi_0\epsi}\sum_{i}\frac{1}{|\mathbf r_i-\mathbf r^\mathrm d_i|},
\end{equation}
and perturbation term
\begin{equation}\label{Eq:DHamI}
\hat H^\mathrm{(1)}_\mathrm{e}=-\frac{e^2}{4\pi\epsi_0\epsi}\left(\frac{1}{|\mathbf r_1-\mathbf r^\mathrm d_2|}+\frac{1}{|\mathbf r_2-\mathbf r^\mathrm d_1|}-\frac{1}{|\mathbf r_1-\mathbf r_2|}\!\right)\!.
\end{equation}
The wave functions of non-interacting electrons are denoted as $\Psi_{1,2}$. 
In the first order perturbation theory the Hamiltonian in Eq.~(\ref{Eq:DHamI}) 
produces the spin-dependent interaction between carriers
\begin{equation}\label{Eq:DHamI1}
\hat H^\mathrm{HL}=4H^{\mathrm{HL}}(\hat{\mathbf s}_1\hat{\mathbf s}_2)=H^{\mathrm{HL}}\cos(\theta).
\end{equation}
Here we introduce the angle $\theta$ between magnetic moments of polarons. Since the polaron magnetic moment is large 
we can treat it as classical value. As was shown in the 
previous section the average spin of electron is co-directed with corresponding polaron magnetic moment. 
The exact formulas for the exchange constant $H^\mathrm{HL}$ is given elsewhere~[\onlinecite{Auerbach}]. 
The only important thing 
for us is that it exponentially decays with the distance between 
donor centres $R$ as $\exp(-2R/\ab)$ and is inversely proportional to $\epsi$. Thus, we can write
\begin{equation}\label{Eq:HLExConst1}
H^\mathrm{HL}=H^\mathrm{HL}_0\frac{e^{-2R/\ab}}{\epsi}.
\end{equation}
Generally, the constant $H^\mathrm{HL}_0$ can be found numerically for wave functions given by Eq.~(\ref{Eq:DWF1}).

\subsection{Superexchange}

Magnetic interaction between two electrons appears also due to virtual hopping 
of electrons between defect sites, so-called superexchange. The coupling 
appears in the second order perturbation theory with respect to the hopping matrix elements, $t=\langle\Psi_1\Psi_1|\hat H_\mathrm{e}|\Psi_1\Psi_2\rangle$. Effective Hamiltonian describing the superexchange is given by~[\onlinecite{Auerbach}]
\begin{equation}\label{Eq:SEHam}
\hat H^\mathrm{se}=\frac{4t^2}{U}(\hat{\mathbf s}_1\hat{\mathbf s}_2).
\end{equation}   
Here $U$ is the onsite repulsion of electrons calculating as $U=\langle\Psi_1\Psi_1|\hat H_\mathrm e|\Psi_1\Psi_1\rangle$. We assume that $U$ is mostly determined by the Coulomb interaction between two electrons situated at the same site. 
$U$ is inversely proportional to the size of the Bohr 
radius and the system dielectric constant, $U\sim (1/(\epsi\ab))$. Hopping matrix element decreases with increasing of distance between the defects, $t^2\sim \exp(-2R/\ab)$. Finally, we arrive to the following  expression for the interaction 
energy
\begin{equation}\label{Eq:SEHam}
\begin{split}
\hat H^\mathrm{se}&=4H^\mathrm{se}_0\ab \epsi e^{-2R/\ab}(\hat{\mathbf s}_1\hat{\mathbf s}_2)=
\\&=H^\mathrm{se}_0\ab \epsi e^{-2R/\ab}\cos(\theta)=H^\mathrm{se}\cos(\theta).
\end{split}
\end{equation}

\subsection{Impurities mediated interaction}

\begin{figure}
	\includegraphics[width=1\columnwidth]{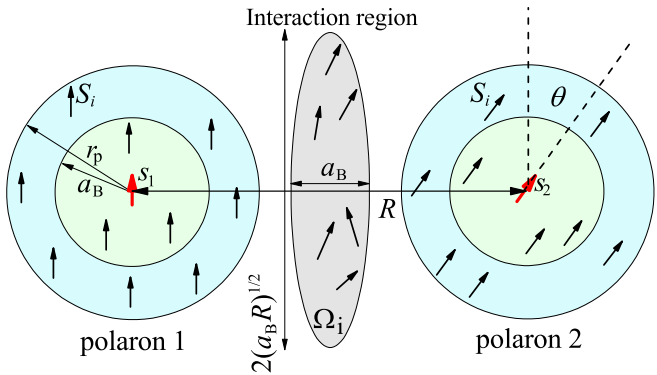}
	\caption{(Color online) Two electro-magnetic polarons in DMFE separated by a distance $R$. Red arrows show direction of electrons average magnetic moments. Angle between magnetic moments of two such electrons is $\theta$. The electrons wave function characteristic size is $\ab$. Black arrows show magnetic moments of impurities in DMFE. Within the magnetic polaron radius $r_\mathrm p$ they are co-directed with average impurity magnetic moment. In the central area between the polarons there is a lens shaped interaction region $\Omega_\mathrm i$. Impurities spins in this region are not fully polarized by donor electrons but correlated with them leading to interaction between the carriers. Width of the interaction region is about $\ab$. Lateral size of the region is about $2\sqrt{R\ab}$. }\label{Fig:TwoPolarons}
\end{figure}

Consider the situation where the distance between polarons $R$ exceeds the single polaron size $r_\mathrm p$ (see Fig.~\ref{Fig:TwoPolarons}). Beyond the polaron radius $r_\mathrm p$ the interaction between electron and impurities much weaker than inside the polaron. In the central region between two polarons impurities interact with both electrons leading to magnetic interaction between carriers (see Fig.~\ref{Fig:TwoPolarons}). We will follow the simplified approach of Ref.~[\onlinecite{Sarma2002}] to calculate this coupling. According to Ref.~[\onlinecite{Sarma2002}] the main contribution to the inter-polaron interaction is given by lens-shaped region with lateral size of $\sqrt{R\ab}$ and width of $\ab$. We assume that interaction of electron with impurities in this regions is independent of impurity position. Magnetic energy of this region is given by 

\begin{equation}\label{Eq:MHamInt}
\hat H^{\mathrm{p-p}}_\mathrm{e-imp}= 2Je^{-R/a_\mathrm B}(\hat{\mathbf{s}}_1+\hat{\mathbf{s}}_2)\sum_{j\in\Omega_\mathrm i}\hat{\mathbf{S}}_j/S_0,
\end{equation}
where summation is over the region of interaction $\Omega_\mathrm i$. Number of impurities inside the interaction region can be estimated as $N_\mathrm i=\pi a_\mathrm B^2 R n_\mathrm i$. Treating the total polarons spins as classical magnetic moments we obtain
\begin{equation}\label{Eq:MHamInt1}
\hat H^{\mathrm{p-p}}_\mathrm{e-imp}= 2Je^{-R/a_\mathrm B}\cos (\theta/2) \sum_{j\in\Omega_\mathrm i}\hat{\mathbf{S}}^{(z)}_j/S_0,
\end{equation}
where $\theta$ is the angle between average magnetic moments of polarons. We assume that both polarons are similar and have the same magnetic moment. $\hat{\mathbf{S}}^{(z)}_j$ is the projection of the impurity spin on the direction $\mathbf m_1+\mathbf m_2$. Interaction of donor electron and impurities in the region $\Omega_\mathrm i$ is weak. 
Therefore, the average magnetic moment created by this interaction is defined as $\langle\sum_{j\in\Omega_\mathrm i}\hat{\mathbf{S}}^{(z)}_j\rangle\approx 2N_\mathrm i S_0 Je^{-R/a_\mathrm B}\cos (\theta/2)/(3k_\mathrm B T)$.  Introducing this result into Eq.~(\ref{Eq:MHamInt1}) we get the average interaction energy of two polarons 
\begin{equation}\label{Eq:MHamInt2}
H^{\mathrm{p-p}}_\mathrm{e-imp}=\frac{4N_\mathrm i J^2e^{-2R/a_\mathrm B}\cos^2 (\theta/2)}{3k_\mathrm B T}=H^{\mathrm{p-p}}(\cos(\theta)+1).
\end{equation}

\section{Magnetic phase diagram of DMFE}\label{Sec:Results}

The distance at which two polarons can be considered as coupled ($r_\mathrm{c}$) is defined by the condition
\begin{equation}\label{Eq:Dist}
|H^\mathrm{HL}+H^\mathrm{se}+H^{\mathrm{p-p}}|=k_\mathrm B T.
\end{equation}
Note that the Heitler-London coupling, $H^\mathrm{HL}>0$, and the superexchange, $H^\mathrm{se}>0$,
produce antiferromagnetic (AFM) coupling while impurity mediated coupling is FM, $H^\mathrm{p-p}<0$. On one hand the first two interactions decay faster with distance ($e^{-2R/\ab}$) than the third one ($e^{-R/\ab}$). But on the other hand the impurity mediated interaction depends on concentration $n_\mathrm i$ and temperature. It decreases with increasing of temperature and reducing of $n_\mathrm i$. Experimental results on DMFE show that in most cases FM order appears at low magnetic impurities concentration~[\onlinecite{Liu2009,Nan2009,Chatterjee2017,Chu2009,Wilson2003,Du2011}] meaning that impurity mediated coupling dominates. However, AFM order is also reported in DMFEs with low impurities concentration~[\onlinecite{Jedrecy2015}].

In the case of $H^\mathrm{HL}_0=H^\mathrm{se}_0=0$, Eq.~(\ref{Eq:Dist}) for the interaction distance $r_\mathrm c$ at given temperature $T$  turns into  
\begin{equation}\label{Eq:FMTrans2}
k_\mathrm B T=\frac{S_0J_0\sqrt{n_\mathrm i r_\mathrm c}e^{-r_\mathrm c/\ab}}{\sqrt{3\pi}a_\mathrm B^2}.
\end{equation}

Approximately one can write
\begin{equation}\label{Eq:IntRad}
r_\mathrm c\sim\ab\left(\!\mathrm{ln}\!\left[\!\frac{S_0J_0}{2\pi\abc k_\mathrm B T}\right]\!+\frac{1}{2}\mathrm{ln}\left[\!\abc n_\mathrm i\mathrm{ln}\!\left(\frac{S_0J_0}{2\pi\abc k_\mathrm BT}\!\!\right)\!\!\right]\!\!\right).
\end{equation}

According to percolation theory~[\onlinecite{Efros}] the long range magnetic order in the system of randomly situated polarons appears approximately at $r_\mathrm c n_\mathrm d^{1/3}=0.86$, where $n_\mathrm d$ is the defects concentration. Introducing $r_\mathrm c$ from this relation into Eq.~(\ref{Eq:Dist})  one can find the ordering temperature. Depending on the sign of the total interaction the ordering can be either FM or AFM (or superspin glass state).

First, consider the case when the polaron-polaron interaction is the dominant one 
and we can neglect the Heitler-London and superexchange contributions. 
In this case there is only FM type interaction between impurities and only the FM/paramagnetic (PM) transition is possible. The transition temperature is given by the equation

\begin{equation}\label{Eq:FMTrans1}
k_\mathrm B T=\frac{S_0J_0\sqrt{0.86 n_\mathrm i}\exp{(-0.86/(\ab n_\mathrm{d}^{1/3}))}}{\sqrt{3\pi}n_\mathrm d^{1/6}a_\mathrm B^2}.
\end{equation}

Note than according to Eqs.~(\ref{Eq:DielT}), (\ref{Eq:Diel}) and (\ref{Eq:DecLength}) the Bohr
radius, $\ab(T,E)$ depends on temperature and external electric field. 
This makes the PM/FM transition temperature more complicated function of $n_\mathrm i$ and $n_\mathrm d$ 
and makes it dependent on electric field.

Dimensionless magnetization of the DMFE is given by the 
following equation~[\onlinecite{Efros,Sarma2002}] 
\begin{equation}\label{Eq:Magnetization}
M(T)=S_0 n_\mathrm i V_\mathrm{inf}((r_\mathrm c(T))^3n_\mathrm d ),
\end{equation}
where $V_\mathrm{inf}$ is the relative volume of infinite cluster (or probability that an impurity belongs to an infinite cluster) in site percolation  problem. We found the function using Monte-Carlo simulations approach developed in Ref.~[\onlinecite{Rerh1969}].

\begin{figure}
	\includegraphics[width=1\columnwidth]{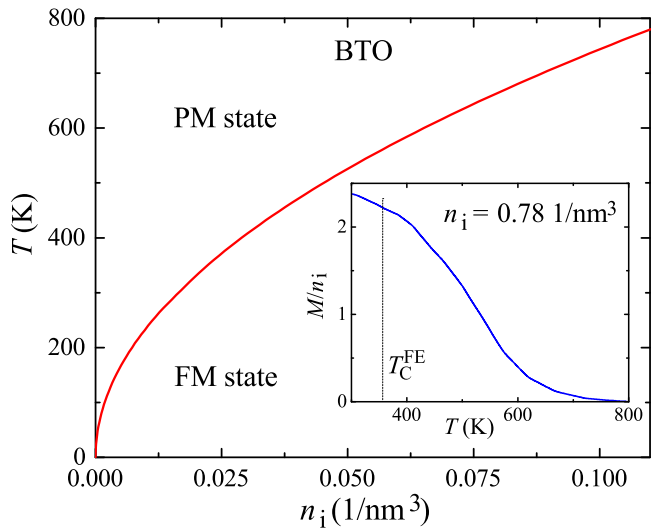}
	\caption{(Color online) Approximate magnetic phase transition temperature Eq.~(\ref{Eq:FMTrans1}) as a function of impurities concentration $n_\mathrm i$. The system parameters, which correspond to BTO doped with Fe, are in the text. 
	The inset is the magnetic moment per Fe impurity as a function of temperature for $n_\mathrm i=0.78$ 1/nm$^3$ ($5$\% doping).}\label{Fig:MPD_BTO}
\end{figure}

\subsection{BaTiO$_3$ based DMFE}

Figure~\ref{Fig:MPD_BTO} shows the magnetic phase diagram of DMFE with the following parameters. Impurities magnetic moment is $S_0=5/2$.  High frequency dielectric constant is $\epsi_{\infty}=5.8$ and the static one is $\epsi=1000$.  
We chose such a value of $m^*$ that $\ab=0.45$ nm. This corresponds to BTO crystal with Fe impurities. Concentration of defects (oxygen vacancies) is about 0.043 nm$^{-3}$ (0.27\%, lattice period in BTO is about $0.4$ nm). Parameter $S_0 J_0=6\cdot 10^4$ K$\cdot$nm$^3$. At impurities concentration of about 5\% this gives spin splitting of the carrier of about 2.4 eV. This splitting occurs due to interaction with all impurities within the polaron. We neglect the 
Heitler-London and superexchange contributions.

The figure shows magnetic state of the system as a function of impurities concentration and temperature.  
The system is FM at low temperatures and high impurities concentration and is PM 
at high temperatures and low concentration of magnetic impurities. 
The curve in Fig.~\ref{Fig:MPD_BTO} shows approximate boundary between these two magnetic states. For such a high dielectric constant the Bohr radius $\ab$ is independent of $\epsi$ and the temperature 
dependence of the dielectric constant does not play any role in magnetic properties of the material.

The inset shows magnetization as a function of temperature for impurities concentration $n_\mathrm i=0.78$ 1/nm$^3$ (5\% for BTO crystal). Magnetic phase transition appears at $T\approx 650$ K. This is in agreement with experiment in Ref.~[\onlinecite{Liu2009}]. Ferroelectric phase transition in BTO appears around $T^\mathrm{FE}_\mathrm C=$360 K. In this region the dielectric constant has a 
strong peak. However, because of very large $\epsi$ the ME effect is weak and
no peculiarities appear in the vicinity of $T^\mathrm{FE}_\mathrm C$.
\begin{figure}
	\includegraphics[width=1\columnwidth]{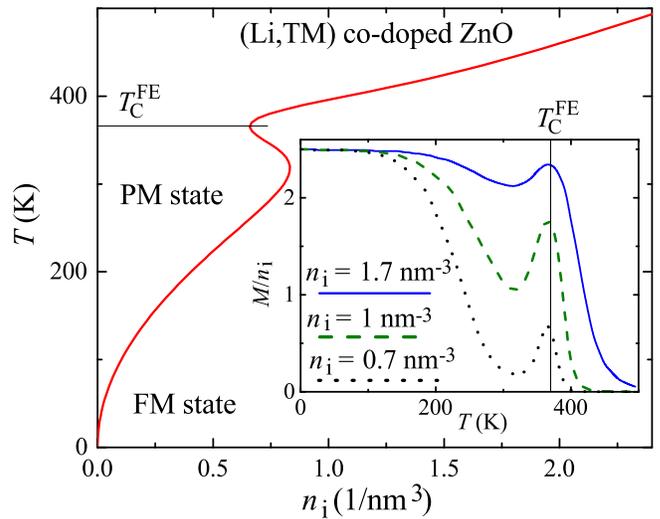}
	\caption{(Color online) Magnetic phase diagram of (Li,TM) co-doped ZnO. The system parameters are provided in the text. The inset is the magnetic moment per TM impurity  as a function of temperature for $n_\mathrm i=1.7$  nm$^{-3}$ (solid blue line), $1$  nm$^{-3}$ (dashed green line) and $0.7$ nm$^{-3}$ (dotted black line).}\label{Fig:MPD_ZnO}
\end{figure}

\subsection{ZnO based DMFE}

(Li,TM) co-doped Zinc oxide is one of the most studied doped magnetic ferroelectrics~[\onlinecite{Kang2010,Kang2011,Nan2007}]. Ordinarily, both ferroelectricity and magnetism in these materials appear due to doping. In contrast to ``classical'' FEs such as BTO, the ZnO based multiferroics have relatively low dielectric constant. Inevitable defects in doped ZnO materials also provides shallow donor states. Due to low dielectric constant of the material the Bohr radius of these states can be temperature dependent.

Magnetism in TM doped ZnO was theoretically and experimentally studied in numerous works~[\onlinecite{Zeng2008,Coey2004,Gehring2008,Waghmare2005}]. Two distinct cases were recognized when the material is either diluted magnetic semiconductor (DMS) or diluted magnetic insulator (DMI)~[\onlinecite{Gehring2008}]. In the first case carriers are delocalized on the scale of the whole sample and magnetic ordering appears due to Ruderman-Kittel-Kasuya-Yosida interaction. In the second case carriers are strongly localized and the coupling is due to magnetic polarons.  We will assume the small concentration of defects and BMP based coupling.

Dielectric and magnetic properties in ZnO-based materials strongly depend on the dopand type, concentration and fabrication procedure. Figure~\ref{Fig:MPD_ZnO} shows magnetic phase digram of the DMFE with parameters close to (TM,Li) co-doped ZnO. Impurities magnetic moment is $S_0=5/2$ and $S_0 J_0=3.3\cdot 10^4$ K$\cdot$nm$^3$ giving the  spin splitting of the electron of about 1.7 eV for impurities concentration $n_\mathrm i=1$ nm$^{-3}$.  High frequency dielectric constant is $\epsi_{\infty}=4$~[\onlinecite{Fitzgerald2005}]. Static dielectric constant strongly depends on temperature with $\epsi^T_\mathrm{min}=25$, $\Delta \epsi^T=45$ and the ferroelectric Curie temperature, $T^\mathrm{FE}_\mathrm{C}=370$ K~[\onlinecite{Zeng2009}]. We chose $m^*$ such that $\ab=0.75$ nm at zero temperature~[\onlinecite{Fitzgerald2005}]. Concentration of defects (oxygen vacancies) is about 0.02 nm$^{-3}$ ($\sim$0.1\%). We neglect Heitler-London and superexchange contributions.

\begin{figure}
	\includegraphics[width=1\columnwidth]{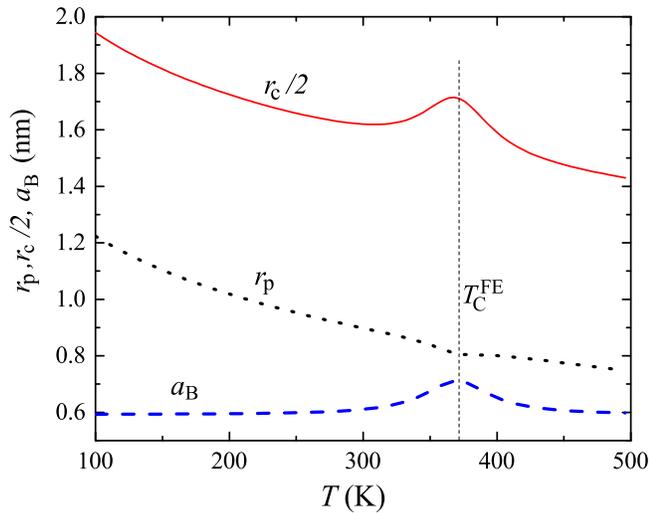}
	\caption{(Color online) The Bohr radius $\ab$, BMP radius $r_\mathrm p$ and BMP interaction distance $r_\mathrm c$ as a function of temperature for DMFE with $\epsi$ described by Eq.~(\ref{Eq:Diel}) with $\epsi_\mathrm{min}=25$, $\Delta\epsi=45$, $T^\mathrm{FE}_\mathrm C=370$ K and $\epsi_{\infty}=4$. Effective mass is chosen 
	such that the Bohr radius at zero temperature is $\ab=0.75$ nm.}\label{Fig:Radius}
\end{figure}

Magnetic phase transition curve has a peculiarity in the vicinity of FE phase transition temperature $T^\mathrm{FE}_\mathrm C=370$ K. The peculiarity is related to non-monotonic behavior of the BMP coupling radius $r_\mathrm c$ in the vicinity of $T^\mathrm{FE}_\mathrm C$ (see Fig.~\ref{Fig:Radius}). Since static dielectric constant is comparable to optical one and it has a peak as a function of temperature at $T=T^\mathrm{FE}_\mathrm C$ the Bohr radius also has a peak in this region. Increasing $\ab$ leads to the increase of BMP interaction distance $r_\mathrm c$ and enhancement of magnetic properties. Note that while the Bohr radius $\ab$ and interaction distance $r_\mathrm c$ have a peak in the vicinity of FE phase transition, the BMP radius $r_\mathrm p$ has a deep (at least for 
given parameters).

Inset in Fig.~\ref{Fig:MPD_ZnO} shows magnetization of DMFE as a function of temperature for several concentrations of TM impurities. Magnetization also has a peak at $T=T^\mathrm{FE}_\mathrm C$. Such a peak is the consequence of coupling between electric and magnetic subsystems in this material and can be considered as magneto-electric effect.

In Refs.~[\onlinecite{Kang2011,Zeng2009}] temperature dependence of (Li,TM) co-doped ZnO magnetization were studied in the vicinity of FE phase transition. No peculiarities in magnetization in the vicinity of the FE transition point were observed. Two possible reasons for the absence of magneto-electric coupling in these particular samples may exist. The first one is that 
samples studied in Ref.~[\onlinecite{Kang2011}] are nanorods of (Li,Co) co-doped ZnO with very large surface/volume ratio. The origin of 
magnetism in such structures is also under question. 
On one hand the conductivity of these samples is small meaning that the material is DMI with possible BMP-based magnetism. On the other hand the magnetism can be related to surface effects as often happens in nanoscale metal oxides~[\onlinecite{Escamilla2011,Rao2009}].  
The second possible reason is that the model of electric polaron described in Sec.~\ref{Sec:SingleElPol} is not applicable to this particular material. Ferroelectricity in this material is related to Li doping and oxygen vacancies~[\onlinecite{Kang2011}]. Electric dipoles in this material are inhomogeneously spread across the sample. Therefore, the low-frequency dielectric constant related to these dipoles should be also rather inhomogeneous. Inhomogeneity of dielectric constant probably appears at the same spatial scale as the distance between magnetic polarons in the system. Therefore, electric dipoles responsible for ferroelectricity and static dielectric constant do not influence the polaron size.

\begin{figure}
	\includegraphics[width=1\columnwidth]{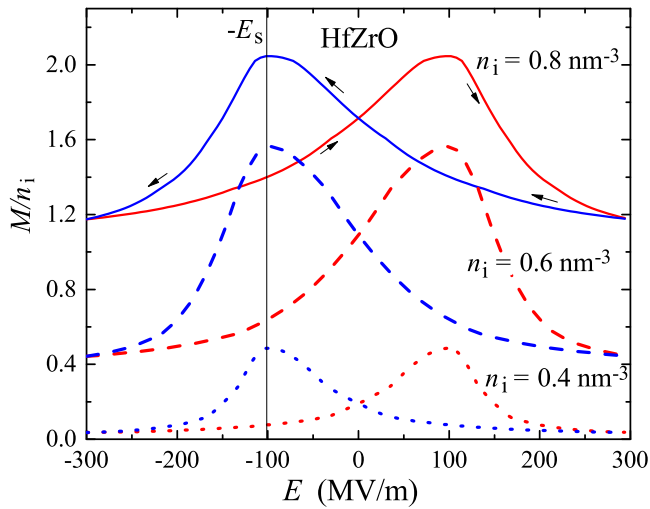}
	\caption{(Color online) Dimensionless magnetic moment per magnetic impurity (maximum moment is $S_0=5/2$) as a function of applied electric field $E$ for $n_\mathrm i=0.8$  nm$^{-3}$ (solid lines), $0.6$  nm$^{-3}$ (dashed lines) and $0.4$ nm$^{-3}$ (dotted  lines). Arrows show the hysteresis bypass direction.}\label{Fig:MvsE_HZO}
\end{figure}

\subsection{Hf$_x$Zr$_{1-x}$O$_2$ based DMFE}

Another FEs family with low dielectric constant is materials based on HfO$_2$. Doping of Hf oxide with various elements leads to the 
appearance of FE properties (spontaneous electric polarization, hysteresis loop, electric field dependent dielectric constant)~[\onlinecite{Mikolajick2011,Frey2011,Mikolajick2012_1}]. In the present work we discuss Hf$_{0.5}$Zr$_{0.5}$O$_2$ FE~[\onlinecite{Mikolajick2012_1}]. This material is homogeneous in contrast to FEs based on weakly doped zinc oxide. This allows to expect that variation of dielectric constant in this material leads to variation of polaron size. The source of carriers in this material is also oxygen vacancies. No data is available on magnetic doping of this material. 

Dielectric constant of  Hf$_{0.5}$Zr$_{0.5}$O$_2$ depends on external electric field. Therefore, one can control magnetic properties of Hf$_{0.5}$Zr$_{0.5}$O$_2$  doped with magnetic impurities using electric field. Figure~\ref{Fig:MvsE_HZO} shows the dependence of magnetization of DMFE on external electric field at room temperature and for different impurity concentrations. Other parameters are chosen as follows. The Bohr radius at zero electric field is 0.5 nm. Impurities magnetic moment is $S_0=5/2$ and $S_0 J_0=3.3\cdot 10^4$ K$\cdot$nm$^3$ giving the  spin splitting of the electron of about 1.1 eV for impurities concentration $n_\mathrm i=1$ nm$^{-3}$.  Defects concentration is $n_\mathrm d=0.05$ nm$^{-3}$ ($\sim 0.6$\% in the case of Hf$_{0.5}$Zr$_{0.5}$O$_2$ which has the lattice constant of 0.5 nm). Optical dielectric constant $\epsi_{\infty}=4.5$. Static dielectric constant as a function of electric field is given by Eq.~(\ref{Eq:Diel}) with $\epsi^E_\mathrm{min}=30$, $\Delta\epsi^E=15$, $E_\mathrm s=100$ MV/m and switching region width $\Delta E_\mathrm s=E_\mathrm s$~[\onlinecite{Mikolajick2012_1}]. The Heitler-London and superexchange contributions are neglected. 

Static dielectric constant depends on electric field leading to electric field dependence of magnetization in the system (ME effect). $\epsi$ demonstrates hysteresis behavior causing hysteresis of magnetization as a function of electric field $E$. Dielectric constant reaches its maximum at the switching field $\pm E_\mathrm s$. According to Eq.~(\ref{Eq:IntRad}) the BMP interaction distance grows with $\epsi$. 
Therefore, the magnetization has peaks at $E=\pm E_\mathrm s$. While interaction distance variation is not large (about 10\%) the magnetization variation is significant.

\subsection{Influence of Heitler-London and superexchange contributions}

Since the Heitler-Lodon and superexchange interactions are antiferromagnetic ones, they compete with the BMP-based coupling. These interactions decay faster with distance between defects than the impurities mediated magnetic coupling, but they do not depend on concentration $n_\mathrm i$ and temperature. Therefore, at low impurities concentration and high temperature antiferromagnetic interactions can dominate leading to antiferromagnetic (or spin glass) ordering. At temperature independent dielectric constant the AFM ordering temperature can be found as follows
\begin{equation}\label{Eq:AFMordTemp}
T^\mathrm{AFM}=\frac{H^\mathrm{HL}+H^\mathrm{se}\pm\sqrt{(H^\mathrm{HL}+H^\mathrm{se})^2+4\tilde H^{\mathrm{p-p}}}}{2},
\end{equation}
where $\tilde H^{\mathrm{p-p}}=-H^{\mathrm{p-p}}k_\mathrm B T$. Solutions exist only if $H^\mathrm{HL}+H^\mathrm{se}>2\sqrt{H^{\mathrm{p-p}}k_\mathrm B T}$. This condition is alway satisfied at low enough impurities concentration. Competition between AFM and FM interactions in DMS was 
considered in Ref.~[\onlinecite{Wolff2002}].

\begin{figure}
	\includegraphics[width=1\columnwidth]{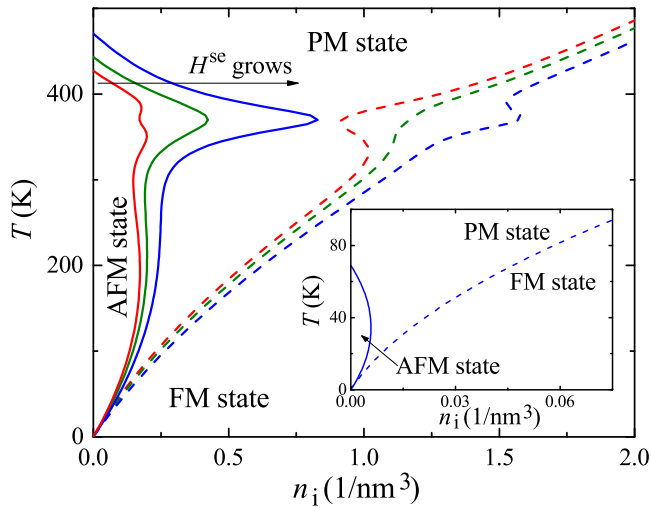}
	\caption{(Color online) Magnetic phase diagram of DMFE with strong Heitler-London and superexchange interactions. Solid lines show boundary between PM and AFM states, dashed lines demonstrate PM/FM transition. All curves are plotted for the same $H^\mathrm{HL}_0$. Red lines correspond to the system with the smallest superexchange contribution, green lines show magnetic phase diagram for a system with intermediate $H^\mathrm{se}_0$ and blue curves are for the highest  $H^\mathrm{se}_0$. All system parameters are provided in the text. 
	Main figure shows the situation of strong direct coupling leading to AFM ordering in the vicinity of FE phase transition.  Inset shows the case when direct coupling ($H^\mathrm{HL}$ and $H^\mathrm{se}$) is weak and induce AFM states only at low temperature.}\label{Fig:MPD_AFM}
\end{figure}

Figure \ref{Fig:MPD_AFM} shows magnetic phase diagram of DMFE with significant contribution of the Heitler-London and superexchange interactions. The following parameters are used. $S_0=5/2$, $S_0 J_0=4\cdot10^4$ K/nm$^3$, $n_\mathrm d=0.02$ nm$^{-3}$, $m^*$ is chosen such that the Bohr radius away 
from $T^\mathrm{FE}_\mathrm C$ is about $0.75$ nm, $\epsi_{\mathrm{\infty}}=4$, $\epsi_{\mathrm{min}}^T=25$, $\Delta \epsi^T=35$, $\Delta T=20$ K, $T^\mathrm{FE}_\mathrm C=370$ K, $H^\mathrm{HL}_0=3.5\cdot10^7$ K (main graph), $H^\mathrm{se}_0=5\cdot10^3$ K/nm (red curves), $10\cdot10^3$ K/nm (green curves), $15\cdot10^3$ K/nm (blue curves). In the inset we use $H^\mathrm{HL}_0=5\cdot 10^6$ K, $H^\mathrm{se}_0=3\cdot10^3$ K/nm.

In contrast to the previously considered cases the region of AFM ordering appears at finite $H^\mathrm{HL}$ and/or $H^\mathrm{se}$. FM/PM boundary also changes. Region of AFM ordering exist only at low impurity concentration, since only in this case AFM interactions overcome strong impurity mediated FM coupling. The main figure shows the case where direct interactions ($H^\mathrm{HL}$ and $H^\mathrm{se}$) are strong and induce AFM ordering at high temperatures close to FE phase transition. Note that Heitler-London interaction decreases with increasing of dielectric constant while the superexchange interaction behaves oppositely. Therefore, behavior of the phase boundaries strongly depends on the ratio between these two contributions. Superexchange mostly influences the region in the vicinity of FE phase transition. AFM region grows and FM region decreases with increasing of $H^\mathrm{se}_0$ in the vicinity of $T^\mathrm{FE}_\mathrm C$. Heitler-London interaction influences the phase diagram aside of $T^\mathrm{FE}_\mathrm C$, but this influence is mostly quantitative. 

Inset shows the case when   $H^\mathrm{HL}$ and/or $H^\mathrm{se}$ are relatively small and do not lead to magnetic ordering in the vicinity of FE phase transition. In this case modifications of FM/PM boundary is weak. AFM region exists 
at low temperatures and low impurities concentration.

\section{Conclusion}

In the present work we proposed a coupling mechanism of magnetic and electric degrees of freedom in doped magnetic ferroelectrics. Magnetic order in DMFE appears due to formation and interaction of BMPs. There are three different contributions into interaction between magnetic polarons. All these contributions depend on dielectric constant of the FE matrix. The most significant is the impurities mediated interaction between polarons. It depends on the radius of polaron wave function. Due to interaction with phonons this radius linearly depends on the dielectric constant of FE matrix. Since the dielectric constant of FEs can be controlled with applied field or varying temperature, one can control the interpolaron interaction and magnetic state of the whole system. Peculiarity of this magneto-electric effect is that it does not involve the relativistic spin-orbit coupling and relies only on the Coulomb interaction.

We calculated magnetic phase transition temperatures as a function of impurities concentration and showed that strong temperature dependence of dielectric permittivity in the vicinity of FE phase transition leads to essential modification of magnetic phase diagram. We found magnetization as a function of temperature and showed that it has a peak in the vicinity of FE phase transition. This peak is a consequence of ME effect appearing in DMFE.

We calculated magnetization as a function of electric field in DMFE and demonstrated that magnetic moment of the system can be effectively controlled with applied bias. Magnetization shows hysteresis behavior as a function of electric field. It has two peaks associated with FE polarization switching.

Strong magneto-electric coupling can appear only in DMFE with low dielectric constant such as (Li,TM) co-doped ZnO or Hf$_{0.5}$Zr$_{0.5}$O$_2$. TM-doped BaTiO$_3$ is not a very promising candidate to observe our effect due 
to very larger dielectric constant.

\section{Acknowledgements}

This research was supported by NSF under Cooperative Agreement Award EEC-1160504 
and NSF PREM Award. O.U. was supported by Russian Science Foundation (Grant  16-12-10340).

\bibliography{DMFE}

\end{document}